\documentclass[12pt]{article}
\usepackage{stmaryrd}
\usepackage{amsfonts}
\usepackage{amsmath}
\usepackage{amssymb}
\usepackage{amscd}
\usepackage{color}

 \linethickness{0.4pt}

\begin{document}
\title{Twisting and $\kappa$-Poincar\'{e}}
\date{}
\maketitle
\begin{center}
{{\large \textbf{$\mathrm{Andrzej\;Borowiec}$, $\mathrm{Jerzy\; Lukierski}$}, }}\\
\bigskip
{Institute of Theoretical Physics\\
University of Wroc{\l }aw pl. Maxa Borna 9, 50-206 Wroc{\l }aw, Poland}\\
e-mail: andrzej.borowiec@ift.uni.wroc.pl; jerzy.lukierski@ift.uni.wroc.pl\\
\bigskip
{{\large \textbf{$\mathrm{Anna\;Pacho{\l}}$}}}\\
\bigskip
 {Science Institute, University of Iceland, Dunhaga 3, 107
Reykjavik, Iceland}\\
e-mail: pachol@hi.is
\end{center}
\bigskip
\begin{abstract}
We demonstrate that the coproduct of D=2 and D=4 quantum $\kappa$-Poincar\'{e} algebra in classical algebra basis can not be obtained by the cochain twist
depending only on Poincar\'{e} algebra generators. We also argue that nonexistence of such a twist does not imply the nonexistence of universal R-matrix. 
\end{abstract}

\section{Introduction}

In this note we would like to present the result that the quantum $%
\kappa $-deformation of Poincar\'{e}-Hopf algebra \cite{1,2,3} written in undeformed Poincar\'{e} algebra basis \cite{8} can not be
obtained by twisting of classical Poincar\'{e}-Hopf algebra even if we relax
the coassociativity condition. We shall consider below the quantum deformations in the
category of triangular quasi-Hopf (tqH) algebras \cite{4,5,5D} with nontrivial
coassociator $\phi \neq 1\otimes 1\otimes 1$. Our arguments are based on the existence of an explicit deformation map called quantum map \cite{8,6,7,9} which connects the generators of $\kappa $-deformed (e.g.
in bicrossproduct basis \cite{3}) and the classical (undeformed) Poincar\'{e}
algebras. In such classical basis the $\kappa$-deformed quantum Poincar\'{e} algebra can be described by classical Poincar\'{e} generators with quite complicated coproducts. Our aim is to show that one can not obtain such $\kappa$-deformed coproducts by a general cochain twist transformation of primitive (classical) Poincar\'{e} coproducts. This result is consistent with the fact that till now the universal R-matrix for $\kappa $%
-deformed Poincar\'{e} algebra is not known, at least in compact form. We argue however (see Appendix) that for D=2 the $\kappa$-deformation the existence of universal R-matrix does not imply the existence of cochain twist defining $\kappa$-deformed coproducts.

It should be added that several
authors \cite{10,11,12,13,14,Dom,Dobrev} have tried  to obtain $\kappa $-deformed Poincar%
\'{e} algebra in the framework of Hopf algebras by proposing nonstandard twists which were embedded in some
enlargements of Poincar\'{e} algebra but they never reproduced in algebraic
way neither the complete $\kappa $-Poincar\'{e} coalgebra sector nor the correct universal R-matrix. We also mention here
 that in recent paper \cite{Stjepan} the twists were studied in the Hopf algebroid framework (see e.g. \cite{Lu,Bohm}) with the use of
 Hopf algebroid structure of deformed phase space but such approach is outside of the standard (quasi)Hopf algebra scheme used in this paper.

The description of quantum deformation by twist provides the universal
R-matrix as well as the explicit formula for the star-product realization of
quantum algebra (see e.g. \cite{17,18}) consistent with the Hopf-algebraic
actions. Unfortunately the standard Poincar\'{e} twist depending only on
Poincar\'{e} generators and satisfying the 2-cocycle condition (i.e. with $%
\phi =1\otimes 1\otimes 1$) cannot describe the quantum $\kappa $%
-deformation of Poincar\'{e} algebra. This property can be deduced also from
the formula for classical $\kappa $-Poincar\'{e} r-matrix (see e.g. \cite{3}; $r\in A\otimes A$) :
\begin{equation}
r=\frac{i}{\kappa }P_{i}\wedge N_{i}
\end{equation}%
which satisfies modified Yang-Baxter equation\footnote{We denote by $P_i$ the three-momenta and $N_i$ are the boost generators.}.

It is known that in the case of deformed semi-simple Lie algebras the canonical
Drinfeld-Jimbo quantum deformation can be obtained by so-called Drinfeld
twist in the category of quasi-Hopf algebras \cite{4,5}. This generalization
of standard quantum groups introduces the category of quantum quasi-groups
which is characterized by the universal R-matrix describing the flip
operation ($(a\otimes b)^{T}=b\otimes a$; $a,b\in A$) on the coproducts
\begin{equation}\label{op}
\Delta ^{T}(a)=R\Delta (a) R^{-1}
\end{equation}
as well as by the non-unital coassociator $\phi =A\otimes A\otimes A$ modifying
the quasi-triangularity relations for the universal R-matrix as follows:
\begin{eqnarray}
\left( \Delta \otimes id\right) (R) &=&
 \phi _{312}R_{13}\phi_{132}^{-1}R_{23}\phi _{123}  \label{DRphi231_b}\\
\left( id \otimes \Delta \right) (R) &=&
\phi _{231}^{-1}R_{13}\phi_{213}R_{12}\phi _{123}^{-1}  \label{DRphi231_a}
\end{eqnarray}

Quasitriangular quasi-Hopf algebras $\tilde{H}=\left( A,\Delta ,S,\epsilon
; R, \phi,\alpha,\beta \right) $ generalize the notion of quasitriangular Hopf
algebras $H=\left( A,\Delta ,S,\epsilon; R \right) $ and are characterized
additionally by coassociator $\phi$ generalizing the coassociativity condition as
follows \footnote{%
The condition (\ref{qcoas}) is called sometimes as describing
quasi-coassociativity, introducing coassociativity up to the similarity transformation.}
\begin{equation}  \label{qcoas}
\left( (id\otimes \Delta )\circ\Delta (a)\right)\,\phi =\phi\,\left( (\Delta \otimes id)\circ\Delta (a)\right)
\end{equation}%
where two special elements $\alpha ,\beta \in A$ are linking antipode $S$
with the coproducts $\Delta$ (see e.g.\cite{4}).

We recall that Poincar\'{e} algebra is not semi-simple, but can be obtained
by Wigner-Inonu contraction of simple AdS or dS algebras. Interestingly
enough, one can introduce suitable quantum modification of Wigner-Inonu
contraction which applied to Drinfeld-Jimbo deformation $U_q(so(3,2))$ of (A)dS algebra provides in the limit $R\shortrightarrow\infty$ (R - AdS radius) the $\kappa $-deformed quantum Poincar\'{e} Hopf algebra \cite{1,19}.
It has been shown however by Young and Zegers \cite{YZ1,YZ2} (see also \cite{Majid-Beggs-0506450,Stjepan}) that
there exists as well the quantum contraction of universal Drinfeld twist for $%
U_{q}(O(3,2))$ (or $U_{q}(O(4,1))$) \footnote{Universal Drinfeld twists in the category of quasi-Hopf algebras are given for quantum semisimple Lie algebras in \cite{4}. They were defined as inner automorphisms which maps by twist the coproducts of quantum and undeformed enveloping algebra. It should be stressed that Drinfeld did prove the existence of universal twist by using cohomological methods, which are not constructive.} what permits to introduce $\kappa $%
-deformed Poincar\'{e}-Hopf algebra as belonging to the category of
triangular quasi-Hopf algebras. 
In this paper we shall perform for particular choice of $\kappa $-Poincar\'{e} coproducts simple
calculations demonstrating that a twist generating the $\kappa $-deformed Poincar\'{e}
algebra from classical (undeformed) Poincar\'{e}-Hopf algebra does not exist
even if one considers general cochain twists allowing noncoassociativity ($\phi \neq 1\otimes 1\otimes 1$)
 \footnote{We stress that the two-cocycle condition is not assumed for the cochain twists.}.

Firstly in Sect. 2 we shall consider simpler case of standard $\kappa $%
-deformed Poincar\'{e} algebra for D=2. Further in Sect. 3 we shall consider
D=4 $\kappa $-deformation. Using perturbation expansion in $\frac{1}{\kappa }
$ deformation parameter it will be shown that the description of coproducts
in terms of twist breaks down already in the second order of the $\frac{1}{%
\kappa }$ power expansion. In order to clarify that our result does not contradict the presence of universal R-matrix for $\kappa$-Poincar\'{e}, which has been calculated perturbatively in \cite{YZ1}, we show in Appendix that in the classical basis the universal R-matrix up to second order of $\frac{1}{\kappa}$ exists (for simplicity we consider D=2 case).

\section{ $\protect\kappa$-Poincar\'{e} from twisting - D=2}

\subsection{ General considerations}

We deal with nontrivial deformations of the enveloping algebra ($U(\hat{g})%
\shortrightarrow U_{q}(\hat{g})$) only if we consider $U_{q}(\hat{g})$ as a
Hopf algebra (or at least as a bialgebra). If we do not consider the
coalgebraic sector there exists always an isomorphism which maps algebraic
sector $U(\hat{g})$ into $U_{q}(\hat{g})$. The quantum deformation,
described infinitesimally by classical r-matrix modifies necessarily only
the coalgebraic sector. Concluding, the basis of the algebraic sector $%
U_{q}(\hat{g})$ of quantum algebra can be chosen classical, but for any
choice of algebra generators the coalgebra will be deformed and described by some
non-symmetric, non-primitive coproducts ($\Delta \neq \Delta ^{T}$).

In the case of $\kappa $-deformed Poincar\'{e} enveloping algebra $U_{\kappa
}(\hat{g})$ $(\hat{g}=(P_{\mu },M_{\mu \nu })\in \mathcal{P}^{3,1})$ one can also show explicitly
the existence of quantum map by presenting the formula transforming
undeformed algebra $U(\hat{g})$ into the deformed one $U_{\kappa }(\hat{g})$%
. First calculations in \cite{6,7} were provided for inverse quantum
map from $\kappa$-deformed basis of Poincar\'{e} algebra to the classical
one, but the explicit calculation of $\kappa -$deformed coproducts of Poincar%
\'{e} generators in classical basis has not been provided. The explicit
formulae for such coproducts were given later in \cite{8,9}.

In this paper firstly we shall discuss the two-dimensional case (D=2) and further the
dimension D=4. We shall assume that the coproducts of $\kappa$-deformed
quantum algebra in classical algebra basis are generated by some twist $F\in
U(\hat{g})\otimes U(\hat{g})$. We expand the logarithm of twist into
the power series in $\frac{1}{\kappa }$ as follows
\begin{equation}
F=\exp \left( \frac{1}{\kappa }f_{1}+\frac{1}{\kappa ^{2}}f_{2}+O\left(
\frac{1}{\kappa ^{3}}\right) \right)  = 1\otimes 1+\frac{1}{\kappa }F_{1}+\frac{1}{\kappa ^{2}}F_{2}+O\left(
\frac{1}{\kappa ^{3}}\right)
\label{twist_exp}
\end{equation}%
where $F_1=f_1$, $F_2={1\over 2}f_1^2+f_2$, etc. This implies the following perturbative formula for the $\kappa$-deformed coproducts $\Delta\in U(\hat{g})\otimes U(\hat{g})[[\frac{1}{\kappa}]]$
\begin{equation*}
\Delta =F\Delta _{0}F^{-1}=\Delta _{0}+\frac{1}{\kappa }\Delta _{1}+\frac{1}{%
\kappa ^{2}}\Delta _{2}+O\left( \frac{1}{\kappa ^{3}}\right) =
\end{equation*}
\begin{equation}
=\Delta _{0}+%
\frac{1}{\kappa }\left[ f_{1},\Delta _{0}\right] +\frac{1}{2\kappa ^{2}}%
\left[ f_{1},\left[ f_{1},\Delta _{0}\right] \right]
+\frac{1}{\kappa ^{2}}\left[ f_{2},\Delta _{0}\right] +O\left( \frac{1}{%
\kappa ^{3}}\right)\label{cop_exp}
\end{equation}
In general case we obtain from $F$ the formula for the universal R-matrix
\begin{equation}\label{Rtwist}
R=F^TF^{-1}
\end{equation}
as well as the nontrivial coassociator $\phi \in
U(\hat{g})\otimes U(\hat{g})\otimes U(\hat{g})[[\frac{1}{\kappa}]]$
(see e.g. \cite{Majid-Beggs-0506450})
\begin{equation}
\phi =(1\otimes F)(id\otimes \Delta )(F)(\Delta \otimes id)(F^{-1})(F^{-1}\otimes
1)  \label{associator}
\end{equation}
One can show that the universal $R$-matrix describing quasi-triangular
quasi-Hopf algebra (see (\ref{DRphi231_a}, \ref{DRphi231_b})) satisfies the
following modified quantum Yang-Baxter equation
\begin{equation}
R_{12}\phi _{312}R_{13}\phi _{132}^{-1}R_{23}\phi _{123}=\phi
_{321}R_{23}\phi ^{-1}_{231}R_{13}\phi _{213}R_{12}
\end{equation}

\subsection{D=2 $\protect\kappa -$Poincar\'{e} algebra in bicrossproduct
basis}

i) algebra
\begin{equation}  \label{bicross}
\left[ \mathcal{P}_{0},\mathcal{P}_{1}\right] =0\quad ,\quad %
\lbrack N,\mathcal{P}_{0}]=i\mathcal{P}_{1}\quad ,\quad \lbrack N,\mathcal{P}%
_{1}]=\frac{i}{2}\kappa \left( 1-\exp \left( -\frac{2\mathcal{P}_{0}}{%
\kappa }\right) \right) -\frac{i}{2\kappa }\mathcal{P}_{1}^{2}
\end{equation}
ii) coalgebra
\begin{eqnarray}
\Delta \left( \mathcal{P}_{0}\right) &=&\mathcal{P}_{0}\otimes 1+1\otimes
\mathcal{P}_{0}; \\
\Delta \left( \mathcal{P}_{1}\right) &=&\mathcal{P}_{1}\otimes 1+\exp \left(
-\frac{\mathcal{P}_{0}}{\kappa }\right) \otimes \mathcal{P}_{1} \\
\Delta \left( N\right) &=&N\otimes 1+\exp \left( -\frac{\mathcal{P}_{0}}{%
\kappa }\right) \otimes N
\end{eqnarray}

\subsection{ D=2 $\protect\kappa -$Poincar\'{e} Hopf algebra in
classical basis}

i) algebra

The D=2 classical Poincar\'{e} algebra described by two-momentum generators $%
P_{\mu }=(P_{0},P_{1})$ and boost generator $N$
\begin{equation}
\left[ P_{0},P_{1}\right] =0\quad ,\quad
\lbrack N,P_{0}]=iP_{1}\quad ,\quad \lbrack N,P_{1}]=iP_{0} \label{cl_P_alg}
\end{equation}%
is derived from $\kappa-$deformed Poicar\'{e} algebra (\ref{bicross}) by the following inverse quantum map
\begin{equation}
P_{0}=\frac{\kappa }{2}\left( \exp \left( \frac{\mathcal{P}_{0}}{\kappa }%
\right) -\exp \left( -\frac{\mathcal{P}_{0}}{\kappa }\right) (1-\frac{1}{%
\kappa ^{2}}\,P_1^{2})\right) \quad ,\quad P_{1}=\mathcal{P}_{1}\exp
\left( \frac{\mathcal{P}_{0}}{\kappa }\right)  \label{inv_qm}
\end{equation}%
Quantum map which is inverse to (\ref{inv_qm}) has the form
\begin{equation}
\mathcal{P}_{0}=\kappa \ln \Pi _{0}\quad ,\quad \mathcal{P}_{1}=P_{1}\Pi
_{0}^{-1}
\end{equation}%
where
\begin{equation}
\Pi _{0}=\frac{1}{\kappa }P_{0}+\sqrt{1-\frac{1}{\kappa ^{2}}C_{0}}\quad
,\quad \Pi _{0}^{-1}=\frac{\sqrt{1-\frac{1}{\kappa ^{2}}C_{0}}-\frac{1}{%
\kappa }P_{0}}{1-\frac{1}{\kappa ^{2}}\,P_1^{2}}  \label{Pi1}
\end{equation}
with $C_{0}$ describing the standard undeformed mass Casimir
\begin{equation}
C_{0}=P^{0}P_{0}+P^{1}P_{1}=-P_{0}^{2}+P_{1}^{2}
\end{equation}
%
ii) coalgebra
\begin{eqnarray}
\Delta \left( P_{0}\right) &=&P_{0}\otimes \Pi _{0}+\Pi _{0}^{-1}\otimes
P_{0}+\frac{1}{\kappa }P_{1}\Pi _{0}^{-1}\otimes P_{1}  \label{cpP_0} \\
\Delta \left( P_{1}\right) &=&P_{1}\otimes \Pi _{0}+1\otimes P_{1} \\
\Delta \left( N\right) &=&N\otimes 1+\Pi _{0}^{-1}\otimes N  \label{cpM_0}
\end{eqnarray}
The $\kappa -$deformed mass Casimir is the following
\begin{equation}
C=\kappa ^{2}\left( \Pi _{0}+\Pi _{0}^{-1}-2-\frac{1}{\kappa ^{2}}%
P_{1}^{2}\Pi _{0}^{-1}\right)
\end{equation}
It can be checked by explicite calculation that the relations (\ref{cpP_0}-\ref{cpM_0}) satisfy D=2
classical Poincar\'{e} algebra (\ref{cl_P_alg}).

\subsection{ No-go theorem for D=2}

One can expand the coproducts (\ref{cpP_0}-\ref{cpM_0}) in powers of $\frac{1%
}{\kappa }$ using
\begin{equation}
\Pi _{0}=1+\frac{1}{\kappa }P_{0}-\frac{1}{2\kappa ^{2}}C_{0}+O\left( \frac{1%
}{\kappa ^{3}}\right) \quad ,\quad \Pi _{0}^{-1}=1-\frac{1}{\kappa }P_{0}+%
\frac{1}{\kappa ^{2}}\left( P_{0}^{2}+\frac{1}{2}C_{0}\right) +O\left( \frac{%
1}{\kappa ^{3}}\right)
\end{equation}
One gets
\begin{equation}
\Delta\left( P_{0}\right) =P_{0}\otimes 1+1\otimes P_{0}+\frac{1}{%
\kappa }P_{1}\otimes P_{1}+
\label{cpP_0_order}
\end{equation}%
\begin{equation}\nonumber
+\frac{1}{\kappa ^{2}}\left( P_{0}^{2}\otimes
P_{0}+\frac{1}{2}C_{0}\otimes P_{0}-\frac{1}{2}P_{0}\otimes
C_{0}-P_{1}P_{0}\otimes P_{1}\right) +O\left( \frac{1}{\kappa ^{3}}\right) ;
\end{equation}
\begin{equation}
\Delta\left( P_{1}\right) =P_{1}\otimes 1+1\otimes P_{1}+\frac{1}{%
\kappa }P_{1}\otimes P_{0}-\frac{1}{2\kappa ^{2}}P_{1}\otimes C_{0}+O\left(
\frac{1}{\kappa ^{3}}\right) ;  \label{cpP_1_order}
\end{equation}%
\begin{equation}
\Delta\left( N\right) =N\otimes 1+1\otimes N-\frac{1}{\kappa }%
P_{0}\otimes N+\frac{1}{\kappa ^{2}}\left( P_{0}^{2}\otimes N+\frac{1}{2}%
C_{0}\otimes N\right) +O\left( \frac{1}{\kappa ^{3}}\right)
\label{cpM_0_order}
\end{equation}
From $\Delta _{1}=\left[ f_{1},\Delta _{0}\right] $ and (\ref{cpM_0_order})
one can easily calculate that \footnote{In fact, $f_1$ is defined up to term $f_1^{(0)}$ for which $[f_1^{(0)},\Delta_0]=0$, such term does not modify our result.}:
\begin{equation}
f_{1}=-iP_{1}\otimes N  \label{f_1}
\end{equation}%
i.e. we get 'half' of classical r-matrix because $r=f_{1}-f_{1}^{T}$.
The equation determining the term $f_{2}$ in (\ref{twist_exp}) looks as
follows:
\begin{equation}
\Delta _{2}=\frac{1}{2}\left[ f_{1},\left[ f_{1},\Delta _{0}\right] \right] +%
\left[ f_{2},\Delta _{0}\right]  \label{f_2}
\end{equation}%
where $\Delta _{2}$ are given explicitly by formulae (\ref{cpP_0_order}-\ref%
{cpM_0_order}). We shall show below that it does not exists such $f_{2}$
which provides $\frac{1}{\kappa ^{2}}$ terms in the coproducts (\ref%
{cpP_0_order}-\ref{cpM_0_order}).

Let us notice that due to (\ref{f_1}) we get $\left[ f_{1},\left[
f_{1},\Delta _{0}\left( N\right) \right] \right] =0$ and we see from (\ref{cpM_0_order}) that $\Delta _{2}\left(
N\right) $ contains in left factors of tensor product the terms quadratic in $P$
and in right ones the terms linear in $N$. Such property due to (\ref{f_2})
implies that if $f_{2}=A_\alpha\otimes B_\alpha$ the factors $A_\alpha$ have to be quadratic in momenta
and factors $B_\alpha$ linear in $N$. In such
circumstances the most general ansatz for $f_{2}$ is the following:
\begin{equation}\label{f2}
f_{2}=\alpha P_{0}^{2}\otimes N+\beta P_{1}^{2}\otimes N+\gamma
P_{0}P_{1}\otimes N+f_2^{(0)}
\end{equation}%
where $[f_2^{(0)},\Delta_0(N)]=0$. We get
\begin{eqnarray}
\Delta _{2}\left( N\right) &=&\alpha \left[ P_{0}^{2},N\right] \otimes
N+\beta \left[ P_{1}^{2},N\right] \otimes N+\gamma \left[ P_{0}P_{1},N\right]
\otimes N= \\
&=&-i\left( \gamma \left( P_{0}P_{0}+P_{1}P_{1}\right) +\left( 2\alpha
+2\beta \right) P_{0}P_{1}\right) \otimes N\nonumber
\end{eqnarray}%
Comparing this result with the coproduct (\ref{cpM_0_order}) we obtain: $%
-i\gamma =\frac{1}{2};\alpha +\beta =0$ what implies that $f_{2}=\frac{i}{2}%
 P_{0}P_{1}\otimes N +\beta C_0\otimes N+f_2^{(0)}$. Further it is easy to see
that for such a choice of $f_{2}$ the terms $P_{0}\otimes C_{0}$ in (%
\ref{cpP_0_order}) and $P_{1}\otimes C_0$ in (\ref{cpP_1_order}) cannot be
obtained from the formula (\ref{f_2}) with any choice of $f_2^{(0)}$ \footnote{The necessity of adding to (\ref{f2}) the nontrivial term $f_2^{(0)}$ was pointed out by V. N. Tolstoy. However it can be shown that such terms does not modify the conclusion obtained if $f_2^{(0)}=0$.
The most important terms belonging to $f_2^{(0)}$ are given by the formula
$f_2^{(0)}=\delta(P_1\otimes NP_0-P_0\otimes NP_1)+\rho(P_1\otimes NP_1 -P_0 \otimes NP_0)$.}. In particular, the term $P_{0}\otimes P^2_{0}$ can never be obtained from the commutator
$[f_2, P_0\otimes 1+ 1\otimes P_0]$ for any element
$f_2\in  U(\hat{g})\otimes U(\hat{g})$. \footnote{A similar argument will be used later for the case $D=4$.}
Concluding we see that the coproducts (\ref{cpP_0}-\ref{cpM_0}) cannot be
obtained by twist and formula (\ref{cop_exp}).

\section{$\protect\kappa $-Poincar\'{e} from twisting - D=4}

\subsection{D=4 $\protect\kappa -$Poincar\'{e} algebra in the
bicrossproduct basis}

i) algebra with Lorentz generators $M_{\mu \nu }=\left(M_{k}=%
\frac{1}{2}\epsilon _{ijk}M_{ij}; N_{i}=M_{0i}\right) $ and four-momenta $\mathcal{P}_\mu=(\mathcal{P}_0,\mathcal{P}_j)$
\begin{eqnarray}
\lbrack M_{i},M_{j}] &=&i\epsilon _{ijk}M_{k}\quad ,\quad \left[ M_{i},N_{j}%
\right] =i\epsilon _{ijk}N_{k}\quad ,\quad \left[ N_{i},N_{j}\right]
=-i\epsilon _{ijk}M_{k}  \label{Poincare_d4_a} \\
\left[ M_{j},\mathcal{P}_{k}\right] &=&i\epsilon _{jki}\mathcal{P}_{i}\quad
,\quad \left[ M_{j},\mathcal{P}_{0}\right] =0\quad ,\quad \left[ N_{j},%
\mathcal{P}_{0}\right] =i\mathcal{P}_{j}  \label{Poincare_d4_b} \\
\left[ N_{i},\mathcal{P}_{j}\right] &=&\frac{i}{2}\delta _{ij}\left[ \kappa
\left( 1-e^{-\frac{2\mathcal{P}_{0}}{\kappa }}\right) +\frac{1}{\kappa }%
\mathcal{\vec{P}}^{2}\right] -\frac{i}{\kappa }\mathcal{P}_{i}\mathcal{P}_{j}
\label{Poincare_d4_c}
\end{eqnarray}
ii) coalgebra

The coproducts satisfying the relations (\ref{Poincare_d4_a}-\ref%
{Poincare_d4_c}) have the form
\begin{eqnarray}
\Delta \left( \mathcal{P}_{0}\right) &=&1\otimes \mathcal{P}_{0}+\mathcal{P}%
_{0}\otimes 1\quad ,\quad \Delta \left( M_{i}\right) =1\otimes
M_{i}+M_{i}\otimes 1  \label{kP1} \\
\Delta \left( \mathcal{P}_{k}\right) &=&e^{-\frac{\mathcal{P}_{0}}{\kappa }%
}\otimes \mathcal{P}_{k}+\mathcal{P}_{k}\otimes 1\quad , \\
\Delta \left( N_{i}\right) &=&N_{i}\otimes 1+e^{-\frac{\mathcal{P}_{0}}{%
\kappa }}\otimes N_{i}-\frac{1}{\kappa }\mathcal{P}_{j}\otimes \epsilon
_{ijk}M_{k}\quad
\end{eqnarray}

\subsection{ D=4 $\protect\kappa $-Poincar\'{e} algebra in classical
basis}
i) algebra

The relations (\ref{Poincare_d4_a}-\ref{Poincare_d4_c}) are mapped into
classical Poincar\'{e} algebra
\begin{eqnarray}
\lbrack M_{i},M_{j}] &=&i\epsilon _{ijk}M_{k}\quad ,\quad \left[ M_{i},N_{j}%
\right] =i\epsilon _{ijk}N_{k}\quad ,\quad \left[ N_{i},N_{j}\right]
=-i\epsilon _{ijk}M_{k} \\
\left[ M_{j},P_{k}\right]  &=&i\epsilon _{jki}P_{i}\quad ,\quad \left[
M_{j},P_{0}\right] =0\quad ,\quad \left[ N_{j},P_{0}\right] =iP_{j}\quad
,\quad \left[ N_{i},P_{j}\right] =i\delta _{ij}P_{0}
\end{eqnarray}%
with the use of the  inverse quantum map:
\begin{equation}
\mathcal{P}_{0}=\kappa \ln \Pi _{0}\quad ,\quad \mathcal{P}_{i}=P_{i}\Pi
_{0}^{-1}
\end{equation}%
where $\Pi _{0}$ and $\Pi _{0}^{-1}$ are given by the formulae (\ref{Pi1})
with four-dimensional classical mass Casimir $%
C_{0}=P^{0}P_{0}+P^{k}P_{k}=-P_{0}^{2}+\vec{P}^{2}$.\\
ii) coalgebra
\begin{eqnarray}
\Delta \left( P_{0}\right) &=&P_{0}\otimes \Pi _{0}+\Pi _{0}^{-1}\otimes
P_{0}+\frac{1}{\kappa }P_{k}\Pi _{0}^{-1}\otimes P_{k} \\
\Delta \left( P_{k}\right) &=&P_{k}\otimes \Pi _{0}+1\otimes P_{k} \\
\Delta \left( M_{i}\right) &=&M_{i}\otimes 1+1\otimes M_{i} \\
\Delta \left( N_{i}\right) &=&N_{i}\otimes 1+\Pi _{0}^{-1}\otimes N_{i}-%
\frac{1}{\kappa }\epsilon _{ikj}P_{k}\Pi _{0}^{-1}\otimes M_{j}
\end{eqnarray}
where $\Pi _{0}$ and $\Pi _{0}^{-1}$ are expanded in ${1\over\kappa}$ power series by means of the formula (17),
with four-dimensional classical mass Casimir $C_{0}$.
By expanding of (39, 40, 42) in $\frac{1}{\kappa}$ we get%
\begin{equation}
\Delta \left( P_{0}\right) =P_{0}\otimes 1+1\otimes P_{0}+\frac{1}{\kappa }%
P_{k}\otimes P_{k}+  \label{cp_P_0_4d}
\end{equation}%
\begin{equation*}
+\frac{1}{\kappa ^{2}}\left( P_{0}^{2}\otimes P_{0}+\frac{1}{2}C_{0}\otimes
P_{0}-P_{k}P_{0}\otimes P_{k}-\frac{1}{2}P_{0}\otimes C_{0}\right) +O\left(
\frac{1}{\kappa ^{3}}\right)
\end{equation*}%
\begin{equation}
\Delta \left( P_{k}\right) =P_{k}\otimes 1+1\otimes P_{k}+\frac{1}{\kappa }%
P_{k}\otimes P_{0}-\frac{1}{2\kappa ^{2}}P_{k}\otimes C_{0}+O\left( \frac{1}{%
\kappa ^{3}}\right)   \label{cp_P_k_4d}
\end{equation}%
\begin{equation}
\Delta \left( N_{i}\right) =N_{i}\otimes 1+1\otimes N_{i}-\frac{1}{\kappa }%
\left( \epsilon _{ikj}P_{k}\otimes M_{j}+P_{0}\otimes N_{i}\right) +
\label{cp_N_i_4d}
\end{equation}%
\begin{equation*}
+\frac{1}{\kappa ^{2}}\left( \left( P_{0}^{2}+\frac{1}{2}C_{0}\right)
\otimes N_{i}+\epsilon _{ikj}P_{k}P_{0}\otimes M_{j}\right) +O\left( \frac{1%
}{\kappa ^{3}}\right)
\end{equation*}%
Following (\ref{f_1}) one can postulate the following formula for D=4 $%
\kappa -$deformation
\begin{equation}
f_{1}=-iP_{i}\otimes N_{i}  \label{f_1_4d}
\end{equation}%
One can check that in accordance with formula (\ref{cop_exp}) one gets
correctly the linear terms in the coproducts (\ref{cp_P_0_4d})-(\ref%
{cp_N_i_4d}).

\subsection{No-go theorem for D=4}
After using formulae (\ref{f_2}) and (\ref{cp_P_0_4d}) we present the term $\Delta _{2}\left( P_{0}\right) $ in two ways:
\begin{eqnarray}
\Delta _{2}\left( P_{0}\right)  &=&-\frac{1}{2}\left[ P_{i}\otimes N_{i},%
\left[ P_{j}\otimes N_{j},P_{0}\otimes 1+1\otimes P_{0}\right] \right] +%
\left[ f_{2},P_{0}\otimes 1+1\otimes P_{0}\right] = \nonumber\\
&=&\frac{1}{2}\vec{P}^{2}\otimes P_{0}+\left[ f_{2},P_{0}\otimes 1+1\otimes
P_{0}\right]  \nonumber  \\
&\overset{?}{=}&\frac{1}{2}\left( P_{0}\otimes P_{0}^{2}+P_{0}^{2}\otimes
P_{0}+\vec{P}^{2}\otimes P_{0}-P_{0}\otimes \vec{P}^{2}\right)
-P_{k}P_{0}\otimes P_{k}\label{no-go}
\end{eqnarray}%
We shall show the impossibility of finding such $f_{2}= A_\alpha\otimes B_\alpha$ that
leads to the validity of last equality in (\ref{no-go}). For that purpose let us
consider only the derivation by twist of the term $P_{0}\otimes P_{0}^{2}$.
One can write the general formula:
\begin{equation}
\left[ f_{2},\Delta _{0}\left( P_{0}\right) \right] =\left[ A_\alpha,P_{0}\right]
\otimes B_\alpha+A_\alpha\otimes \left[ B_\alpha,P_{0}\right]   \label{AB}
\end{equation}%
Because $A_\alpha$ and $B_\alpha$ depend only on Poincar\'{e} generators, the commutators $\left[
A_\alpha,P_{0}\right] $ and $\left[ B_\alpha,P_{0}\right] $ will necessarily generate the
term different from $P_{0}$; in fact because $\left[ P_{i},P_{0}\right] =%
\left[ M_{i},P_{0}\right] =0$ the two commutators on right-hand side of (\ref{AB}) are different
from zero only if $A_\alpha$ and/or $B_\alpha$ depends on boost generators $N_{i}$. Because $O(3)$-covariance of $\kappa$-deformed algebra and coalgebra relations is their basic feature \cite{1,2,3}, we assumed that the term $f_2$ is $O(3)$-invariant, i.e. generators $N_i$ in $f_2$ are always accompanied by $P_i$, on the same or other side of the tensor product $\otimes$. By writing down all possible $O(3)$-invariant terms in $f_2$ linear in Lorentz generators (and quadratic in momenta for dimensional reasons) it can be shown that the term $P_{0}\otimes P_{0}^{2}$ can never be
obtained by twist from the formula (\ref{AB}).

The only way of satisfying the equality (\ref{no-go}) would be to consider the elements $A_\alpha$ and $B_\alpha$ containing some
generators $X$ which satisfy the commutation relation $\left[ X,P_{0}\right] \sim P_{0}$. Such
generator $X$ can be provided only if we enlarge the  Poincar\'{e}
symmetries, in particular by the scale transformations with $X$ identified with the dilatations generator $\mathcal{D}$. In such a way
the twist $F$ will be introduced as spanned by the generators of the eleven-dimensional extension
$\left( P_{\mu },M_{\mu \nu },\mathcal{D}\right) $ of the D=4 Poincar\'{e} algebra called
also D=4 Weyl algebra. It can be mentioned that such enlarged "outer" Poincar\'{e} twists
depending on the dilatation generators $\mathcal{D}$ have been considered before, (see
e.g. \cite{10,14,Stjepan}) and were used in order to describe the star-product providing $\kappa
$-Minkowski noncommutative space-time. In particular in [15] the twist used for the construction of universal
R-matrix depends on the phase space realization of dilatation generator.

\section{Discussion and outlook}
The existence of universal Drinfeld twist generating quantum $\kappa$-Poincar\'{e} algebra from undeformed one by twist \cite{4}-\cite{5D} in category of quasi-Hopf algebras has been proven in \cite{YZ2} by the use of cohomological arguments. The proof was based on the analogy with the quantum (depending on deformation parameter) contraction $R\shortrightarrow\infty$ (R - AdS radius) of $U_q(so(3,2))$ to $\kappa$-deformed Poincar\'{e}
Hopf algebra \cite{1}. But it should be mentioned that\\
- however it is known the explicit formula for universal R-matrix for $U_q(so(3,2))$ (see e.g. \cite{KT}), due to the appearance of divergent terms in the quantum contraction limit which we are not able to compensate, it was not possible to obtain universal R-matrix for $\kappa$-Poincar\'{e} by quantum contraction.\\
- however the cohomological arguments imply the existence of universal Drinfeld twist for D=4 $\kappa$-Poincar\'{e} \cite{YZ2} we do not have even an explicit formula for Drinfeld twist describing the $U_q(o(3,2))$ deformation \footnote{Partial explicit results for Drinfeld twists are known only for $U_q(sl(2;2))$ quantum algebra and its real forms (see e.g. \cite{twistsu2}, \cite{Dabrowski}).} before quantum $\kappa$- contraction.\\

It should be recalled, however, that the universal R-matrix has been calculated perturbatively (in standard basis \cite{YZ1} up to fifth order in $\frac{1}{\kappa}$) as quasi-Hopf algebra with nontrivial coassociator. In order to demonstrate that nonexistence of cochain twist defining $\kappa$-deformed coproducts is not in contradiction with the existence of universal R-matrix we did calculate for D=2 in classical algebra basis (see Appendix) the universal R-matrix up to second order in $\frac{1}{\kappa}$. If we define universal R-matrix by perturbative expansion of its logarithm
\begin{equation}\label{univR}
R=\exp(\frac{1}{\kappa}r_1 +\frac{1}{\kappa^2}r_2+...)
\end{equation}
and calculate $r_1$, $r_2$ using relation (\ref{op}), and the perturbative expansions (\ref{cpP_0_order}-\ref{cpM_0_order}) we get the following formula (see (A.8)) 
\begin{eqnarray}\label{r1}
r_1&=&iP_1\wedge N\\
r_2&=& \frac{i}{2}(N\wedge P_1P_0 + NP_0\wedge P_1)\label{r2}
\end{eqnarray}
We see that the nonexistence of twist satisfying relation (\ref{cop_exp}) does not prohibit the existence of universal R-matrix. In fact, if we express the R-matrix (\ref{univR}-\ref{r2}) by the formula (\ref{Rtwist}), the twist factor $F$ will not satisfy the basic relation  (\ref{cop_exp}).

We would like to point out that in order to consider twist as belonging to the tensor product of classical Poincar\'{e} enveloping algebras $\left(U(\mathcal{P}^{3,1})\otimes U(\mathcal{P}^{3,1})\right)[[\frac{1}{\kappa}]]$ we performed the calculation in particular classical $\kappa$-Poincar\'{e} basis defined by maps (16) and (17).
In order to conclude that $\kappa$-Poincar\'{e} is not endowed with triangular quasi-Hopf algebra structure, i.e. contest the results of \cite{YZ2}, the statement of the nonexistence of twist should be shown for all isomorphisms between the $\kappa$-Poincar\'{e} algebra basis and the undeformed one. Indeed our formulae (\ref{inv_qm}) correspond to particular fixed choice of such isomorphism, but we mention that these formulas were generalized in \cite{7} in a way containing two arbitrary parameters. The generalization of our discussion here to more general choice of the quantization map, e.g. given in \cite{7}, we shall consider in a near future.

Concluding we add also that the problem which can be studied further is the consideration of generalized $\kappa $-deformations (see e.g. \cite{LLM,KMLS}) with the extension of standard $\kappa $-deformed Minkowski space relations by the introduction of arbitrary constant four-vector $\tau ^{\mu }:$
\begin{equation}\label{tkm}
\left[ \hat{x}^{\mu },\hat{x}^{\nu }\right] =\frac{i}{\kappa }\left( \tau
^{\mu }\hat{x}^{\nu }-\tau ^{\nu }\hat{x}^{\mu }\right)
\end{equation}%
Standard $\kappa $-deformed Poincar\'{e} algebras considered in
Sect. 2 and 3 correspond to the choice $\tau ^{\mu }=\left(1,0\right) (D=2)$ and $\tau ^{\mu }=\left( 1,0,0,0\right) (D=4) $%
. The general $\kappa$-deformed Poincar\'{e} algebras which lead to the relations (\ref{tkm}) can be split into two cases:

i) $\tau ^{\mu }\tau _{\mu }=0$ (light-like constant four-vector determining
light-cone $\kappa $-deformation). In such a case following old results from
nineties \cite{Ballesteros,Mudrov} one can show that the twist $F$ providing
the coproducts by formula (\ref{cop_exp}) does exist and satisfies as well
the two-cocycle condition (i.e. generates by formula (\ref{associator}) the
trivial coassociator $\phi =1\otimes 1\otimes 1$).

ii) if $\tau ^{\mu }\tau _{\mu }\neq 0$ one can demonstrate that the result is
Sect. 3.3 about the nonexistence of twist remains valid for any choice of time-like $%
\tau _{\mu }$ (equivalent to standard $\kappa $-deformation) and for any space-like $\tau
_{\mu }$ (tachyonic $\kappa $-deformation).

\section*{Acknowledgements}

Authors would like to thank Valery N. Tolstoy for valuable remarks.
This work is supported by the Polish National Science Centre (NCN) project
2011/01/B/ST2/03354. AP acknowledges the financial support from Erasmus
Mobility and Training Program under the contract no. 2/FSS/DYD/2013/2014 as
well as the hospitality of the Institute for Theoretical Physics, University
of Wroclaw, in November-December 2013.

\section*{Appendix}

\textbf{Perturbative expansion of the universal R-matrix for D=2 $\kappa$-Poincar\'{e} in classical basis.}

The  universal R-matrix describes the flip operation ($(a\otimes
b)^{T}=b\otimes a$; $a,b\in A$) on the coproducts (see also (\ref{op})):
\begin{equation}\nonumber
\Delta ^{T}(a)=R\Delta (a)R^{-1}\qquad\qquad\qquad\qquad\qquad\qquad\qquad\qquad\qquad(A.1)
\end{equation}
We assume (see also \cite{YZ1}) that universal R-matrix can
be expanded as in formula \ref{univR} and one gets $\frac{1}{\kappa}$ expansion of transposed coproducts
and:$\quad $%
\begin{eqnarray}\nonumber
\Delta ^{T}\left( a\right) &=&R\Delta \left( a\right) R^{-1}=\Delta +\left[
r,\Delta \right] +\frac{1}{2}\left[ r,\left[ r,\Delta \right] \right] +...=
\\
&=&\Delta +\frac{1}{\kappa }\left[ r_{1}, \Delta \right] +\frac{1}{2\kappa
^{2}}\left[ r_{1},\left[ r_{1},\Delta \right] \right] +\frac{1}{\kappa ^{2}}%
\left[ r_{2},\Delta \right] +O\left( \frac{1}{\kappa ^{3}}\right)\qquad\qquad\qquad(A.2)\nonumber
\end{eqnarray}
where
\begin{equation}\nonumber
\Delta =\Delta _{0}+\frac{1}{\kappa }\Delta _{1}+\frac{1}{\kappa ^{2}}\Delta
_{2}+O(\frac{1}{\kappa ^{3}})\qquad\qquad\qquad\qquad\qquad\qquad\qquad\qquad(A.3)
\end{equation}
We obtain:
\begin{eqnarray}\nonumber
\Delta ^{T}\left( a\right) &=& \Delta _{0}+\frac{1}{\kappa }\Delta _{1}+\frac{1}{\kappa ^{2}}\Delta _{2}+\frac{1}{\kappa }\left[ r_{1},\Delta
_{0}+\frac{1}{\kappa }\Delta _{1}\right] +\\ \nonumber
&+&\frac{1}{2\kappa ^{2}}\left[ r_{1},\left[ r_{1},\Delta _{0}\right] \right] +\frac{1}{\kappa ^{2}}\left[ r_{2},\Delta _{0}\right]
+O\left( \frac{1}{\kappa ^{3}}\right) \\ \nonumber
&=& \Delta _{0}+\frac{1}{\kappa }\left( \Delta _{1}+\left[ r_{1},\Delta _{0}%
\right] \right) +\qquad\qquad\qquad\qquad\qquad\qquad\qquad(A.4)\\ \nonumber
&+&\frac{1}{\kappa ^{2}}\left( \Delta _{2}+\left[ r_{1},\Delta _{1}\right] +%
\left[ r_{2},\Delta _{0}\right] +\frac{1}{2}\left[ r_{1},\left[ r_{1},\Delta
_{0}\right] \right] \right)+O\left( \frac{1}{\kappa ^{3}}\right) 
\end{eqnarray}
Comparing order by order we get the following equations
\begin{eqnarray}
\Delta _{0}^{T}&=&\Delta _{0}\qquad\qquad\qquad\qquad\qquad\qquad\qquad\qquad\qquad\qquad(A.5a)\nonumber\\
\Delta _{1}^{T}&=&\Delta _{1}+\left[ r_{1},\Delta _{0}\right]\label{cop_op1}\qquad\qquad\qquad\qquad\qquad\qquad\qquad\qquad\qquad(A.5b)\nonumber\\
\Delta _{2}^{T}&=&\Delta _{2}+\left[ r_{1},\Delta _{1}\right]
+\left[ r_{2},\Delta _{0}\right] +\frac{1}{2}\left[ r_{1},\left[
r_{1},\Delta _{0}\right] \right]\label{cop_op2}\qquad\qquad\qquad\qquad\qquad(A.5c)\nonumber
\end{eqnarray}
where $\Delta ^{T}=\Delta _{0}^{T}+\frac{1}{\kappa }\Delta _{1}^{T}+\frac{%
1}{\kappa ^{2}}\Delta _{2}^{T}+O\left( \frac{1}{\kappa ^{3}}\right) $.

The expansion of opposite coproducts in the classical D=2 $\kappa$-Poincar\'{e}
basis are the following:
\begin{equation}
\Delta^{op}\left( P_{0}\right) =P_{0}\otimes 1+1\otimes P_{0}+%
\frac{1}{\kappa }P_{1}\otimes P_{1}+\qquad\qquad\qquad\qquad\qquad\qquad\qquad(A.6a)\nonumber
\end{equation}%
\begin{equation}
+\frac{1}{\kappa ^{2}}\left( P_{0}\otimes P_{0}^{2}+\frac{1}{2}P_{0}\otimes
C_{0}-\frac{1}{2}C_{0}\otimes P_{0}-P_{1}\otimes P_{1}P_{0}\right) +O\left( 
\frac{1}{\kappa ^{3}}\right) ;  \notag
\end{equation}%
\begin{equation}
\Delta^{op}\left( P_{1}\right) =P_{1}\otimes 1+1\otimes P_{1}+%
\frac{1}{\kappa }P_{0}\otimes P_{1}-\frac{1}{2\kappa ^{2}}C_{0}\otimes
P_{1}+O\left( \frac{1}{\kappa ^{3}}\right) ;\qquad\qquad\qquad\qquad\qquad(A.6b)\nonumber
\end{equation}%
\begin{equation}
\Delta^{op}\left( N\right) =N\otimes 1+1\otimes N-\frac{1}{\kappa 
}N\otimes P_{0}+\frac{1}{\kappa ^{2}}\left( N\otimes P_{0}^{2}+\frac{1}{2}%
N\otimes C_{0}\right) +O\left( \frac{1}{\kappa ^{3}}\right)\qquad\qquad\qquad\quad\quad(A.6c)\nonumber
\end{equation}
One can check easily, that $r_{1}=iP_{1}\wedge N$ satisfies eq. (\ref{cop_op1}) for all generators:$N,P_{0},P_{1}$.

Eq. (\ref{cop_op2}) applied to D=2 $\kappa$-Poincar\'{e} gives: 
\begin{equation}\label{r2P0}
\left[ r_{2},\Delta _{0}\left( P_{0}\right) \right] =P_{1}P_{0}\otimes
P_{1}-P_{1}\otimes P_{1}P_{0}\qquad\qquad\qquad\qquad\quad\qquad\qquad\qquad\qquad(A.7a)\nonumber
\end{equation}%
\begin{equation}\label{r2P1}
\left[ r_{2},\Delta _{0}\left( P_{1}\right) \right] =\frac{1}{2}\left(
P_{0}^{2}\otimes P_{1}-P_{1}\otimes P_{0}^{2}-P_{0}\otimes
P_{1}P_{0}+P_{1}P_{0}\otimes P_{0}\right) \qquad\qquad\qquad\qquad\qquad(A.7b)\nonumber
\end{equation}%
\begin{eqnarray}\label{r2N}
\left[ r_{2},\Delta _{0}\left( N\right) \right]  &=&N\otimes P_{0}^{2}+\frac{%
1}{2}N\otimes C_{0}-P_{0}^{2}\otimes N-\frac{1}{2}C_{0}\otimes N-\frac{1}{2}%
NP_{0}\otimes P_{0}\nonumber\\
&&+\frac{1}{2}P_{1}\otimes NP_{1}+\frac{1}{2}P_{0}\otimes P_{0}N-\frac{1}{2}P_{1}N\otimes P_{1}\qquad\qquad\qquad\qquad\qquad(A.7c)\nonumber
\end{eqnarray}
One can show that taking 
\begin{equation}
r_{2}=\frac{i}{2}\left( N\wedge P_{1}P_{0}-NP_{0}\wedge P_{1}\right) \qquad\qquad\qquad\qquad\qquad\qquad\qquad\qquad\qquad(A.8)\nonumber
\end{equation}
the equations (\ref{r2P0}-\ref{r2N}) are satisfied, since
\begin{eqnarray}\label{comm_r2}
&&[N\otimes P_1 P_0 +NP_0\otimes P_1 -P_1 P_0\otimes N -P_1\otimes NP_0, N\otimes 1+1 \otimes N]=\qquad\qquad\qquad(A.9)\nonumber\\
&&=NP_1 \otimes P_1 - P_0^2 \otimes N - P_1^2 \otimes N -P_0 \otimes NP_0 +N\otimes (P_0^2 +P_1^2)+NP_0 \otimes P_0 -P_1\otimes NP_1\nonumber
\end{eqnarray}
The perturbative calculations in D=2 can be extended also to D=4 case.


\begin{thebibliography}{99}
\bibitem{1} J. Lukierski, A. Nowicki, H. Ruegg, V.N. Tolstoy, Phys. Lett. B 264, 331 (1991)

\bibitem{2} J. Lukierski, A. Nowicki, H. Ruegg, Phys. Lett. B 293, 344
(1992);
\bibitem{3} S. Majid, H. Ruegg, Phys. Lett. B 329, 189 (1994);
\bibitem{8} A. Borowiec, A. Pachol, J. Phys. A: Math. Theor. 43, 045203
(2010) [arXiv:0903.5251];
\bibitem{4} V.G. Drinfeld, Algebra Anal. 1, no 6, 114 (1989) (in Russian); Leningr. Math. J. 1, no 6, 1419 (1990);
\bibitem{5} V.G. Drinfeld, Algebra Anal. 1, no 2, 30 (1989) (in Russian);
Leningr. Math. J. 1, no 2, 321 (1990);
\bibitem{5D} V.G. Drinfeld, Algebra Anal. 2, no 4,  149 (1990) (in Russian);
Leningr. Math. J. 2, no 4, 829 (1990);

\bibitem{6} H. Ruegg, V. N. Tolstoy, Lett .Math. Phys 32, 85 (1994);
[arXiv:hep-th/9406146]

\bibitem{7} P. Kosinski, J. Lukierski, P. Maslanka, J. Sobczyk, Mod. Phys.
Lett. A10, 2599 (1995)



\bibitem{9} A. Borowiec, A. Pachol, Eur. Phys. J. C 74, 2812 (2014) [arXiv:1311.4499]

\bibitem{10} A. Ballesteros, N. R. Bruno, F. J. Herranz, Symmetry methods in
Physics, edited by C. Burdik, O. Navratil and S. Posta, Joint Institute for
Nuclear Research, Dubna (Russia), pp. 1-20, (2004) [arXiv:hep-th/0409295]

\bibitem{11} J. G. Bu, H-C. Kim, Y. Lee, C. H. Vac, J. H. Yee, Phys. Lett. B
665, 95 (2008); [arXiv:hep-th/0611175]

\bibitem{12} M. Arzano, A. Marciano, Phys. Rev. D 76, 125005 (2007)

\bibitem{13} T. R. Govindarajan, S. Kumar Gupta, E. Harikumar, S. Meljanac,
D. Meljanac, Phys. Rev. D 77, 105010 (2008); [arXiv:0802.1576]

\bibitem{14} A. Borowiec, A. Pachol, Phys. Rev. D 79, 045012 (2009);
[arXiv:0812.0576]
\bibitem{Dom}D. Kovacevic, S. Meljanac
 J. Phys. A: Math. Theor. 45 (2012) 135208 [arXiv:1110.0944];
\bibitem{Dobrev}  V.K. Dobrev,
J. Phys. A 26, 1317 (1993).
\bibitem{Stjepan} S. Meljanac, A. Samsarov, R. Strajn,
JHEP 08 (2012) 127 [arXiv:1204.4324]
\bibitem{Lu} J.H. Lu, Int. J. Math. 7, 47 (1996)
\bibitem{Bohm} G. Boehm, Handbook of Algebra Vol. 6, ed. M. M. Harzewinkel, Elsevier, 173-236 (2009) [arXiv:0805.3806]

\bibitem{17} C. Blohmann, J. Math. Phys. 44, 4736 (2003);
[arXiv:math/0209180]

\bibitem{18} P. P. Kulish, Proc. of Karlstadt International Conference
(July, 2004), ed. AMS, Contemporary Mathematics 391, 213 (2005);
[arXiv:hep-th/0606056]

\bibitem{19} G. Amelino-Camelia, L. Smolin, A. Starodubtsev, Class. Quant.
Grav. 21, 3095 (2004); [arXiv:hep-th/0306134]

\bibitem{YZ1} C. A. S. Young, R. Zegers, Nucl. Phys. B 804, 342 (2008);
[arXiv: 0803.2659]

\bibitem{YZ2} C. A. S. Young, R. Zegers, Comm. Math. Phys. 298, 585 (2010);
[arXiv: 0812.3257]

\bibitem{Majid-Beggs-0506450} E. J. Beggs, S. Majid, J. Math. Phys. 51,
053522 (2010) [arXiv:math/0506450];
\bibitem{KT} S.M.  Khoroshkin, V.N. Tolstoy, Comm. Math. Phys. 141, 599 (1991);
\bibitem{twistsu2} C. Blohmann, J. Math. Phys. 46,  053519 (2005); [arXiv:math/0410448]
\bibitem{Dabrowski} L. Dabrowski, F. Nesti, P. Siniscalco, Proc. 12th Italian Conf.Gen.Rel.Grav. Phys. Roma 1996; World Sci., Singapore (1997);
[arXiv:q-alg/9610012]


\bibitem{LLM} J. Lukierski, V. D. Lyakhovsky, M. Mozrzymas, Phys. Lett. B
538, 375 (2002); [arXiv:hep-th/0203182]

\bibitem{KMLS} P. Kosinski, P. Maslanka, J. Lukierski, A. Sitarz,
Proceedings of the Conference ''Topics in Mathematical Physics, General
Relativity and Cosmology'', On the occasion of the 75th Birthday of Jerzy F.
Plebanski, 17.09-20.09 2002, Mexico City, Eds: H. Garcia-Compean et.al.,
World Scientific (2003); [arXiv:hep-th/0307038]

\bibitem{Ballesteros} A. Ballesteros, F.J. Herranz, M.A. del Olmo and M.
Santander, Phys. Lett. B 351, 137 (1995);

\bibitem{Mudrov} A. I. Mudrov, "Twisting cocycles in fundamental
representation and triangular bicrossproduct Hopf algebras"
[arXiv:math/9804024]


\end{thebibliography}
\end{document}